\documentclass[showpacs,prb,twocolumn]{revtex4}
\usepackage{amsfonts}

\usepackage{graphics}
\usepackage{epsfig}
\usepackage{graphicx}
\usepackage{dcolumn}
\usepackage{bm}
\usepackage{amssymb,amsmath}
\usepackage{array}

\begin{document}

\title{Absence of Supercritical Behavior in Gapped Graphene with Short-range
Impurity Scattering}
\author{Stepan Grinek$^{1,4}$, Zhou Li$^{2}$, Jie Chen$^{1,4\ast}$, Qinwei
Shi$^{3}$, Frank Marsiglio$^{2}$}
\address{$^1$Electrical and Computer Engineering, University of
Alberta, Alberta, Canada T6G 2V4}
\address{$^2$Department of Physics, University of Alberta, Canada T6G 2V4}
\address{$^3$Hefei National Laboratory for Physical Sciences at
Microscale, University of Science and Technology of China, Hefei
 230026, China}
\address{$^4$National Research Council/National Institute of
Nanotechnology, Alberta, Canada T6G 2M9}\email[Electronic
address:]{jc65@ualberta.ca}
\date{\today}

\begin{abstract}
We show that the changes in the electronic density of states (DOS) in
graphene induced by localized impurities (single, double and multiple) are
significantly different from those caused by the long-range Coulomb
potential. We focus on gapped graphene; a bound state is present within the
gap, with a certain amount of spectral weight. As the coupling to the
impurity increases, the state lowers in energy and approaches the lower
continuum valence band. The spectral weight of this state does not transform
into a resonance state in the valence band, so no unusual screening effects
related to a redistribution of DOS in the continuum is observed. In terms of
the continuous Dirac limit, this phenomenon is a consequence of the absence
of the ``potential bump" at infinity, which is present in the potential of
the effective Schr\"odinger equation for graphene with long-range Coulomb
impurity potential. The states induced by short-range impurity scattering in
graphene, therefore, have distinctly different properties compared with
the long-range potential case. These properties closely resemble the case of
a short-range single impurity in other bipartite lattices, such as square,
body-centered cubic, and simple cubic lattices. For these bipartite
lattices, there is always a localized bound state with energy in the band
gap for the entire range of on-site coupling strengths. In all cases the
energy of these states asymptotically approaches the edge of the valence
band as the magnitude of the coupling strength increases, but never crosses
it.
\end{abstract}

\maketitle


\address{$^1$Electrical and Computer Engineering, University of
Alberta, Alberta, Canada T6G 2V4}
\address{$^2$Department of Physics,
University of Alberta, Canada T6G 2V4}
\address{$^3$Hefei National Laboratory for Physical Sciences at
Microscale, University of Science and Technology of China, Hefei
 230026, China}
\address{$^4$National Research Council/National Institute of
Nanotechnology, Alberta, Canada T6G 2M9}

\section{Introduction}

Graphene has been used as an analog case to study relativistic phenomena in
heavy atoms because the critical charge for the Dirac fermions in graphene
is of the order of unity \cite{Pereira08}. Upon introduction of a single
charge impurity into an otherwise perfect lattice, a bound state appears
within the mass gap that separates  the original Dirac cones \cite%
{ElectronicStruct, Zhou07}. As one tunes the coupling strength of this
charge impurity to larger values, eventually a critical charge condition is
achieved whereby the energy of the bound states passes into the continuum.
In gapped graphene, this crossover has implications for the screening
properties of the electron gas.\cite{Pereira08} The wave function of such a
bound state decays exponentially with the distance from the charge. As the
coupling strength becomes supercritical, the screening effect is
sufficiently strong that the observed effective charge is reduced by almost
4 elementary units compared with the unscreened case.\cite{Pereira08} This
phenomenon is similar to what happens in atomic physics when the elementary
charge is around $170$.\cite{Zeldovich} The shape of the cloud of screening
charge closely follows the shape of the so called ``critical state" just
before it merges with the continuum of the valence band. A couple of
questions arise: i) Is there a change in the electronic structure that
depends on the range of potential? ii) What are the properties that define
whether the impurity potential strength is critical or not?

The paper is organized as follows. In the next section we obtain the
predictions of the long-wavelength Dirac formalism in the case of one or few
short-range impurities. In the third section, after we briefly review the
Green functions on the lattice pertinent to the honeycomb lattice, we obtain
analytical results for one and two impurities on the graphene lattice. In
the fourth section, we generalize our results for multiple short-ranged
impurities, and then we close with the conclusion.

\section{ Dirac equation with spherical well}

The motion of an electron with a fixed full (pseudo-spin plus orbital)
angular momentum in a circularly symmetric potential $U(r)$ is:
\begin{eqnarray}
(E-U(r)-m)A(r)-(\partial _{r}+\frac{j}{r})B(r) &=&0,  \notag \\
(\partial _{r}-\frac{j}{r})A(r)+(E-U(r)+m)B(r) &=&0,  \label{Dirac}
\end{eqnarray}%
where, for definiteness, $U(r)=V\theta (a-r)$, where $\theta (x) $ is a
Heaviside step function, $a$ is the radius of the well, and $V$ is negative
(positive) for a well (barrier). The full wave function is \cite{Pereira08}:
\begin{equation}
\Psi _{C}(r,\phi )=\frac{1}{\sqrt{r}}\left\{
\begin{array}{c}
e^{-i(j-1/2)\phi }A(r) \\
ie^{-i(j+1/2)\phi }B(r)%
\end{array}%
\right\}.  \label{eqn2}
\end{equation}
Here j is an eigenvalue of angular momentum $J_z = L_z+\frac{1}{2}\sigma_z$,
and $C$ can refer to either the $A$ or $B$ sublattice. To solve Eq. (\ref%
{Dirac}), one can express $B(r)$ in terms of $A(r)$:
\begin{equation}  \label{B}
B(r)=\frac{ \frac{j}{r}A(r)-A^{\prime}(r)}{E -V +m}.
\end{equation}

In what follows, we will discuss the case when the energy level is at either
gap edge; for negative (positive) $V$ this is the lower (upper) edge.
The solutions inside the well are:
\begin{eqnarray}
A(r) &=&C_{1}\sqrt{r}\sqrt{\frac{E+m-V}{E-m-V}}J_{j-1/2}(\sqrt{%
(E-V)^{2}-m^{2}}r),  \notag \\
B(r) &=&C_{1}\sqrt{r}J_{j+1/2}(\sqrt{(E-V)^{2}-m^{2}}r).  \label{solI}
\end{eqnarray}%
The solutions outside the well are:
\begin{eqnarray}
A(r) &=&C_{2}\sqrt{r}\sqrt{\frac{E+m}{E-m}}K_{j-1/2}(\sqrt{m^{2}-E^{2}}r),
\notag \\
B(r) &=&-iC_{2}\sqrt{r}K_{j+1/2}(\sqrt{m^{2}-E^{2}}r),
\end{eqnarray}%
where $J_{\alpha}(x)$ is a Bessel function of the first kind, $K_{\alpha}(x)$
is a modified Bessel function of the second kind chosen to satisfy the
boundary condition at infinity. The equation for the energy levels is:
\begin{eqnarray}
\frac{\sqrt{\frac{E+m-V}{E-m-V}}J_{j-1/2}(\sqrt{(E-V)^{2}-m^{2}}a)}{%
J_{j+1/2}(\sqrt{(E-V)^{2}-m^{2}}a)} &=&  \notag  \label{secular} \\
i\frac{\sqrt{\frac{E+m}{E-m}}K_{j-1/2}(\sqrt{m^{2}-E^{2}}a)}{K_{j+1/2}(\sqrt{%
m^{2}-E^{2}}a)} &&.
\end{eqnarray}%
For a solution to exist, the terms on the left-hand side and on the right-hand side 
must be
either pure imaginary or pure real numbers. This means, that for $-m\leq E\leq 0$, and
with $V<0$, 
then we require that $(E-V)^{2}>m^{2}$ for a solution. We are primarily
interested in negative values of $V$.

\begin{figure}[tp]
\begin{center}
\includegraphics[height=3.0in,width=3.0in]{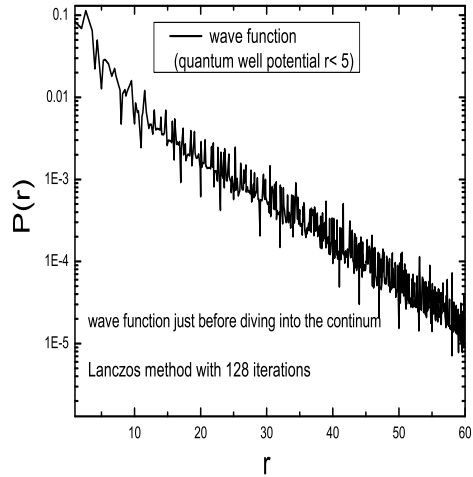}
\end{center}
\caption{Asymptotic behavior of wave function close to the band edge for
quantum well potential.}
\label{fig1}
\end{figure}

We are interested also in what sequence the levels will merge into the
continuum, depending on their angular momentum. One can numerically analyze
Eq. (\ref{secular}) but for the sake of clarity we will analyze the
effective Schrodinger equation for $A(r)$, obtained from the system (\ref%
{Dirac}):

\begin{equation}  \label{Scr1}
-A^{\prime\prime}(r)+[\frac{j^2-j}{r^2}-(E-V)^2+m^2]A(r)= 0
\end{equation}
This is a wave equation for a particle in a potential with functional form $%
\frac{j^2-j}{r^2}-(E-V)^2+m^2$ at zero energy. Obviously, for states with a
higher value of $j^2-j$, the potential curve is higher, particularly near
the origin; states with a higher value of $j$ will merge into the continuum
for larger values of $|V|$. Another way of looking at the properties of the
wave function as these solutions merge into the continuum is to find
asymptotic solutions outside the well when $E \rightarrow- m+0$:
\begin{eqnarray}
|\Psi _{A}|^{2} &=&|A(r)|^{2}/r\simeq \frac{C_{2}^{2}}{r}\frac{E+m}{E-m}%
\frac{1}{\sqrt{m^{2}-E^{2}}}e^{-2\sqrt{m^{2}-E^{2}}r}  \notag \\
|\Psi _{B}|^{2} &=&|B(r)|^{2}/r\simeq -\frac{C_{2}^{2}}{r}\frac{1}{\sqrt{%
m^{2}-E^{2}}}e^{-2\sqrt{m^{2}-E^{2}}r}.  \label{solO_asymp}
\end{eqnarray}%
Note that the wave function disappears on A-sites and $\Psi_B^2$ becomes
non-normalizable when $E =- m$.

As we will see from the following, the Dirac approximation for graphene
gives correct predictions for electronic properties at the critical coupling
strength, because the energies involved are right at the boundary of the
valence band, where the linear dispersion is most accurate.

\section{Analytical results for one and two impurities on the lattice}

\subsection{Lattice Green function}

The Hamiltonian of a free electron in the two-dimensional graphene lattice,
using the tight binding model is,
\begin{equation}
H_{0}=-t\sum_{\mathbf{j,\delta }}(a_{\mathbf{j}}^{\dagger }b_{\mathbf{j}+%
\mathbf{\delta }}+b_{\mathbf{j+\delta }}^{\dagger }a_{\mathbf{j}})+m\sum_{%
\mathbf{i}}(a_{\mathbf{i}}^{\dagger }a_{\mathbf{i}}-b_{\mathbf{i}}^{\dagger
}b_{\mathbf{i}}),
\end{equation}%
where $a_{j}^{\dagger }$ is the creation operator of an electron on the
A-atom site labeled $j$ in the honeycomb lattice, and $b_{j+\delta }$
represents the annihilation of an electron on the neighboring B-atom site
labeled $j+\delta $. Here $\delta $ denotes the three vectors that connect
an A-atom site to its three nearest neighboring B-atom sites. The parameters
$t$ and $m$ represent the nearest neighbor hopping probability and the mass
differentiating the $A$ and $B$ sublattices, respectively.
The Hamiltonian in k-space can be written as:
\begin{equation}
\widehat{H}_{0}=\left(
\begin{array}{cc}
m & \phi _{k} \\
\phi _{k}^{\ast } & -m%
\end{array}%
\right),
\end{equation}%
where we have adopted the standard spinor notation for the $A$ and $B$
sublattice components of the wave function. Here, $\phi
_{k}=-te^{-ik_{x}a}(1+2\cos (k_{y}\sqrt{3}a/2)e^{ik_{x}a3/2})$, where $a$ is
the distance between neighboring atoms. The eigenvalues are
\begin{equation}
\epsilon_{\mathbf{k}, \pm} = \pm \sqrt{t^2(1 + 4c_y^2 + 4c_x c_y) + m^2},
\label{dispersion}
\end{equation}
where $c_x \equiv \cos{3k_xa/2}$ and $c_y \equiv \cos{\sqrt{3}k_ya/2}$. The
Green functions in k-space can be obtained straightforwardly as
\begin{equation}
\mathbb{G}_{0}(\mathbf{k},i\omega _{n})=\left[
\begin{array}{cc}
G_{AA}^{0}(\mathbf{k},i\omega _{n}) & G_{AB}^{0}(\mathbf{k},i\omega _{n}) \\
G_{BA}^{0}(\mathbf{k},i\omega _{n}) & G_{BB}^{0}(\mathbf{k},i\omega _{n})%
\end{array}%
\right],
\end{equation}%
where $i\omega _{n}=i\pi T (2n-1)$ (with $n$ an integer) are the Fermion
Matsubara frequencies ($T$ is the temperature), and each component is given
by
\begin{eqnarray}
G_{AA}^{0}(\mathbf{k},i\omega _{n})&=& \frac{i\omega _{n}+\mu +m}{\left(
i\omega _{n}+\mu \right) ^{2}-\varepsilon_{k}^{2}}  \notag \\
G_{BB}^{0}(\mathbf{k},i\omega _{n})&=& \frac{i\omega _{n}+\mu -m}{\left(
i\omega _{n}+\mu \right) ^{2}-\varepsilon_{k}^{2}}  \notag \\
G_{AB}^{0}(\mathbf{k},i\omega _{n})&=&\frac{\phi _{k}^{\ast }}{\left(
i\omega _{n}+\mu \right)^{2}-\varepsilon_{k}^{2}}  \notag \\
G_{BA}^{0}(\mathbf{k},i\omega _{n})&=&\frac{\phi _{k}}{\left( i\omega
_{n}+\mu \right) ^{2}-\varepsilon _{k}^{2}}.  \label{greenarray}
\end{eqnarray}
We have added $\mu $, the chemical potential, for completeness, and the
superscript `0' serves to remind us that these Green functions are
applicable to the clean lattice, i.e. without impurity scattering. The
lattice Green functions in real space can be obtained by Fourier transform
from the above Green's functions:%
\begin{equation*}
\mathbb{G}_{0}(l,j,i\omega _{n})=\left[
\begin{array}{cc}
G_{AA}^{0}(l,j,i\omega _{n}) & G_{AB}^{0}(l,j,i\omega _{n}) \\
G_{BA}^{0}(l,j,i\omega _{n}) & G_{BB}^{0}(l,j,i\omega _{n})%
\end{array}%
\right],  \label{greenarray_real}
\end{equation*}%
where
\begin{eqnarray}  \label{G_int1}
&&G_{AA}^{0}(l,j,i\omega _{n})=\frac{(\omega +\mu +m)}{4\pi ^{2}}\int_{-\pi
}^{\pi }du\int_{-\pi }^{\pi }dv \\
&&\times \frac{e^{iu(l_{x}-j_{x})}e^{-iv(l_{x}-j_{x})}e^{i2v(l_{y}-j_{y})}}{%
\left( \omega +\mu \right) ^{2}-m^{2}-t^{2}(1+4\cos ^{2}v+4\cos u\cos v)},
\notag
\end{eqnarray}

\begin{eqnarray}
&&G_{AB}^{0}(l,j,i\omega _{n})=-\frac{1}{4\pi ^{2}}\int_{-\pi }^{\pi
}du\int_{-\pi }^{\pi }dv  \label{G_int2} \\
&&\times \frac{%
e^{iu(l_{x}-j_{x})}e^{-iv(l_{x}-j_{x})}e^{i2v(l_{y}-j_{y})}(1+2e^{-iu}\cos v)%
}{\left( \omega +\mu \right) ^{2}-m^{2}-t^{2}(1+4\cos ^{2}v+4\cos \nonumber %
u\cos v)}.
\end{eqnarray}%
The remaining components are readily obtained through the relations $%
G_{BB}^{0}(l,j,i\omega _{n})=G_{AA}^{0}(l,j,i\omega _{n},m\rightarrow -m)$
and, for the off-diagonal components, $G_{BA}^{0}(l,j,i\omega
_{n})=[G_{AB}^{0}(l,j,i\omega _{n})]^{\ast }$. Through the analytic
continuation, $i\omega _{n}\rightarrow \omega +i\delta $, we obtain the
Green functions slightly above the real axis, corresponding to the retarded
Green function. For the particular case of $l=j$ (an on-site Green
function), we can obtain the diagonal components analytically. From now on
we set t=1, which means that all energies are measured in units of the hopping
energy. The result for $G_{AA}^{0}$ is (we set $\mu =0$ for simplicity, and
use the definition $E^{2}=\left\vert \omega ^{2}-m^{2}\right\vert $):

(i) For $0<E<1,\omega ^{2}-m^{2}<0$,
\begin{eqnarray*}
\mathrm{Re}[G_{AA}^{0}(l,l,\omega )]& = &-\frac{(\omega +m)}{\pi }\frac{2}{%
\sqrt{\left( \sqrt{E^{2}+1}\right) ^{3}\sqrt{E^{2}+9}}} \\
&&\times F(\frac{\pi }{2},\frac{1}{2}\sqrt{\frac{-(E^{4}+12E^{2}-6)}{\left(
\sqrt{E^{2}+1}\right) ^{3}\sqrt{E^{2}+9}}+2}), \\
\mathrm{Im}[G_{AA}^{0}(l,l,\omega )] &= & 0.
\end{eqnarray*}%
(ii) For $0<E<1,\omega ^{2}-m^{2}>0$%
\begin{eqnarray*}
\mathrm{Re}[G_{AA}^{0}(l,l,\omega )]&=&-\frac{(\omega +m)}{\pi }\frac{4}{%
\sqrt{3-E}\left[ \sqrt{\left( E+1\right) }\right] ^{3}} \\
&&\times F(\frac{\pi }{2},\sqrt{\frac{\left[ 3+E\right] \left( 1-E\right)
^{3}}{\left[ 3-E\right] \left( E+1\right) ^{3}}}),
\end{eqnarray*}%
\begin{eqnarray*}
\mathrm{Im}[G_{AA}^{0}(l,l,\omega )]&=&-\frac{2(\omega +m)}{\pi }\frac{1}{%
\sqrt{3-E}\left[ \sqrt{\left( E+1\right) }\right] ^{3}} \\
&&\times F(\frac{\pi }{2},\sqrt{\frac{16E}{\left[ 3-E\right] \left(
E+1\right) ^{3}}}).
\end{eqnarray*}%
(iii) for $1<E<3$%
\begin{eqnarray*}
\mathrm{Re}[G_{AA}^{0}(l,l,\omega )] &= & \frac{(\omega +m)}{2\pi }\frac{1}{%
\sqrt{E}} \times F(\frac{\pi }{2},\sqrt{\frac{\left( E+3\right) \left(
E-1\right) ^{3}}{16E}}), \\
\mathrm{Im}[G_{AA}^{0}(l,l,\omega )] &=& -\frac{(\omega +m)}{2\pi }\frac{1}{%
\sqrt{E}}F(\frac{\pi }{2},\sqrt{\frac{\left[ 3-E\right] \left( E+1\right)
^{3}}{16E}}).
\end{eqnarray*}%
(iv) for $E> 3$
\begin{eqnarray*}
\mathrm{Re}[G_{AA}^{0}(l,l,\omega )]&=&\frac{2(\omega +m)}{\pi }\frac{1}{%
\sqrt{\left( E+3\right) \left( E-1\right) ^{3}}} \\
&&\times F(\frac{\pi }{2},\sqrt{\frac{16E}{\left( E+3\right) \left(
E-1\right) ^{3}}}), \\
\mathrm{Im}[G_{AA}^{0}(l,l,\omega )]&=&0.
\end{eqnarray*}
In these expressions we have used $F({\frac{\pi }{2}},x) \equiv
\int_0^{\pi/2} (1 - x \sin^2{\theta})^{-1/2} \ d\theta$, which is a complete
elliptic integral of the first kind (also denoted as $K(x)$).\cite{abram}

\subsection{Analytical formalism}

We start with the equation
\begin{equation}
\hat{G}=(\hat{I}-\hat{G}^0 \hat{V})^{-1}\hat{G}^0,
\end{equation}
where $\hat{G}$ denotes a matrix where different rows (and columns)
correspond to different lattice sites (an explicit example below will make
this clearer, see also Ref. \onlinecite{Callaway}).

As an example, we consider the specific case with two on-site impurities
located at the sites labeled $0$ and $1$ (without loss of generality, we
number the first atom on the $A$ sublattice as $0$, and we number the first
atom on the $B$ sublattice as $1$; since the lattice is bi-partite, $A$%
-atoms are denoted by even numbers, and $B$-atoms are denoted by odd
numbers). The $\hat{I}-\hat{G}_0 \hat{V}$ matrix is then written explicitly
as
\begin{equation}  \label{I-G_0V}
\hat{I}-\hat{G}_0\hat{V} = \begin{pmatrix} 1-VG^0_{00}& -VG^0_{01}& 0 & 0 &
\dots\\ -VG^0_{10}& 1-VG^0_{11}& 0 & 0 & \dots\\ -VG^0_{20}& -VG^0_{21}& 1 &
0 & \dots\\ -VG^0_{30}& -VG^0_{31}& 0 & 1 & \dots\\
\hdotsfor[2]{5}\end{pmatrix},
\end{equation}
where now the subscripts refer to the two site indices (previously written
as arguments in, say, Eq. (\ref{G_int1})), and $V$ is the strength of the
impurity potential at both sites. Then
\begin{equation}  \label{G}
G_{jj}=\sum_k (I-G^0 V)^{-1}_{jk} G^0_{kj}=C_{kj} G^0_{kj}/\Delta,
\end{equation}
where $\mathbb{C}$ is a cofactor in the matrix (\ref{I-G_0V}), and $\Delta$
is the determinant of $\hat{I}-\hat{G}^0 \hat{V} $. The factor $C_{kj}$ is $%
(-1)^{i+j}$ times the determinant of the original matrix excluding the $k$%
-th row and $j$-th column. Eq. (\ref{G}) can be expressed as:
\begin{eqnarray}  \label{G_exp}
&&G_{jj}= C_{kj} G^0_{kj}/\Delta \\
&&=\left[ \sum_k C_{jj} G^0_{jj}+ \sum_{k\leq l, k\neq j} C_{kj} G^0_{kj} +
\sum_{k>l, k\neq j} C_{kj} G^0_{kj} \right] {\large {/} \Delta}.  \notag
\end{eqnarray}
Here the number $l$ is given by the number of sites occupied by an impurity.

For $j>l$ (away from the impurities), $\sum_{k>l, k\neq j} C_{kj} G^0_{kj}=0$
in Eq. (\ref{G_exp}). In this case Eq. (\ref{G_exp}) becomes:
\begin{equation}  \label{G_exp2}
G_{jj}=\sum_k C_{kj} G^0_{kj} {\large {/} \Delta}= \left[ C_{jj} G^0_{jj}+
\sum_{k\leq l, k\neq j} C_{kj} G^0_{kj} \right] {\large {/} \Delta}.
\end{equation}
The cofactor $C_{jj}$ is equal to $\Delta$, and therefore
\begin{equation}  \label{G_exp2.5}
G_{jj}= G^0_{jj}+ \left[ \sum_{k\leq l, k\neq j} C_{kj} G^0_{kj} \right]
{\large {/} \Delta}.
\end{equation}
Note that when $\omega=-m$, $G^0_{00}(\omega+i\delta) \sim 0$.\newline

\subsection{Single impurity scattering}

When there is only one impurity at any $A$-atom site, the Hamiltonian is
given by $\hat{H}=\hat{H}_{0}+\hat{H}_{1},$ where $\hat{H}_{1}= \hat{V}$.
The corresponding Green functions corresponding to the two Hamiltonians are $%
\hat{G}_{0}(z)=(z-\hat{H}_{0})^{-1}$ and $\hat{G}(z)=(z-\hat{H})^{-1}.$ By
using the T-matrix expansion, we obtain the Green function in the presence
of a single impurity:
\begin{equation}
G_{ij}=G^0_{ij}+ {\frac{G^0_{i0} V G^0_{0j} }{1 - V G^0_{00}}}.  \label{gone}
\end{equation}
The local density of states (LDOS) at any position on the graphene lattice
is defined by the imaginary part of Green function:
\begin{equation}
\rho (j,j,\omega )=-\frac{1}{\pi }\mathrm{Im} G_{jj}(\omega +i\delta ),
\end{equation}
and we have restored the explicit frequency dependence for clarity.

The local density of states at the impurity site is $\rho (0,0,\omega )=-%
\frac{1}{\pi }\mathrm{Im}\left( \frac{G_{00}^{0}(\omega+i\delta)}{%
1-VG_{00}^{0}(\omega+i\delta)}\right).$ The position of the bound state in
the gap is determined by the solution of the equation $1-VG_{00}^{0}(%
\omega+i\delta)=0$. At the lower band edge where $\omega \rightarrow -m,$\ $%
G_{00}^{0}(\omega+i\delta)\sim \omega +m=0$. Therefore, inspection of the
above equation suggests that no solution exists, unless $V \rightarrow
-\infty$. This observation implies that for any $V$ the bound state will not
merge into the lower continuum, i.e. no bound state energy crosses the edge
at $\omega = -m$. For a single impurity on a $B$-atom site, the same remarks
apply for a positive impurity potential, and the upper band edge plays the
role previously played by the lower band edge.

To understand how the bound state approaches the continuum band edge, we use
the asymptotic expansion of the complete elliptic integral of the first kind
\cite{abram} to get 
%
%
\begin{equation}
G_{00}(\omega +i\delta)\simeq \frac{K(\omega +m)\ln \left\vert \omega
+m\right\vert }{1-VK(\omega +m)\ln \left\vert \omega +m\right\vert };
\label{lead_G_00}
\end{equation}%
where $K = \frac{1}{ \sqrt{3} \pi}$ By expanding the Green function near $%
\omega=-m$ we obtain a pole with spectral weight $a_0=-K\omega
_{1}^{2}\ln \omega _{1},$ where $\omega _{1}\equiv \omega + m$ is the
solution of%
\begin{equation}
1-VK\omega _{1}\ln \omega _{1}=0.
\end{equation}
It is clear that as $\omega_1 \rightarrow 0$ a solution will only occur as $%
V \rightarrow -\infty$, and the residue corresponding to that solution
approaches zero.

At the other extreme, for a very weak (negative) impurity potential, a
similar expansion near $\omega \sim m$ gives a bound state energy
asymptotically approaching the upper band edge (let $\omega_2 \equiv m -
\omega$):
\begin{equation}
\omega_2 \approx \exp{\frac{-1}{2m K|V|}}.  \label{weak}
\end{equation}

The spectral weight approaches zero here as well, as $a_0 =2 m K \omega_2 \ln^2\omega_2$,
which also goes to zero as the upper band edge is approached.

To summarize the results of this section, we showed that, as the (negative)
impurity potential decreases from zero towards negative infinity, the
frequency of the pole migrates from $+m$ (upper band edge) to $-m$ (lower
band edge). As this occurs, the spectral weight first starts from zero,
grows to some maximum, and then decreases again to zero, as the strength of
the potential varies from zero to negative infinity.

\section{Two or more impurity scattering}

\subsection{Exact solution for two impurities}

We now consider the two-impurity case, with one on an $A$-site, $(0,0)$, and
the second on a $B$-site \textbf{i}(i$_{x}$,i$_{y}$). The Hamiltonian is
\begin{equation*}
\hat{H}=\hat{H}_{0}+\hat{V}+ \hat{V}_{2} = \hat{H}_{1}+\hat{H}_{2},
\label{3.1}
\end{equation*}%
where $\hat{H}_{1} = \hat{H}_{0}+\hat{V}$ as in the single impurity case.
The Green functions $G^{0}$, $G^{1}$ and $G$ correspond to $\hat{H}_{0},\hat{%
H}_{1}$ and $\hat{H}$, respectively. The T-matrix for this case is
\begin{equation*}
\hat{T}=\hat{H}_{2}+\hat{H}_{2}\hat{G}^{1}\hat{H}_{2}+...  \label{3.2}
\end{equation*}%
Therefore, the Green function becomes
\begin{equation*}
G_{jk}=G^1_{jk}+ {\frac{G^1_{j\mathbf{i}} V_2 G^1_{\mathbf{i}k} }{1 - V_2
G^1_{\mathbf{i},\mathbf{i}}}}.  \label{gtwo}
\end{equation*}

In fact, for the many-impurity case, the T-matrix method can be used in a
recursive way,
\begin{equation*}
\hat{G}^{n}=\hat{G}^{n-1}+\hat{G}^{n-1}\hat{T}_{n}\hat{G}^{n-1},  \label{3.4}
\end{equation*}
where%
\begin{equation*}
(T_{n})_{\mathbf{i},\mathbf{i}} = \frac{V_{n}}{1-V_{n}(G^{n-1})_{\mathbf{i,i}%
}}.  \label{3.5}
\end{equation*}

To compute the local density of states at site $(0,0)$, we need $%
G_{00}(\omega )$ (for simplicity we suppress the $i\delta $):
\begin{eqnarray}
&&G_{00}(\omega ) \nonumber \\
&=&G_{00}^{1}(\omega )+\frac{G_{01}^{1}(\omega )V_{2}G_{10}^{1}(\omega )}{%
1-V_{2}G_{11}^{1}(\omega )} \nonumber \\
&=&{\frac{G_{00}^{0}(\omega )\left[ 1-V_{2}G_{11}^{0}(\omega
)\right]
+V_{2}|G_{01}^{0}(\omega )|^{2}}{\left[ 1-V_{2}G_{11}^{0}(\omega )\right] %
\left[ 1-VG_{00}^{0}(\omega )\right] -VV_{2}|G_{01}^{0}(\omega )|^{2}}} \nonumber \\
&=&\frac{F(\omega )}{(\omega -\omega _{0})+i\delta },  \label{3.6}
\end{eqnarray}%
where $\omega _{0}$ is the energy of the pole, and $F(\omega )$ accounts for
the remaining (non-singular) frequency dependence. The actual pole position
is the solution of
\begin{equation}
\left[ 1-V_{2}G_{11}^{0}(\omega _{0})\right] \left[ 1-VG_{00}^{0}(\omega
_{0})\right]
=VV_{2}|G_{01}^{0}(\omega _{0})|^{2},  \label{3.7}
\end{equation}
and the spectral weight is given by $F(\omega _{0})$.

\begin{figure}[tp]
\begin{center}
\includegraphics[height=3.0in,width=3.0in]{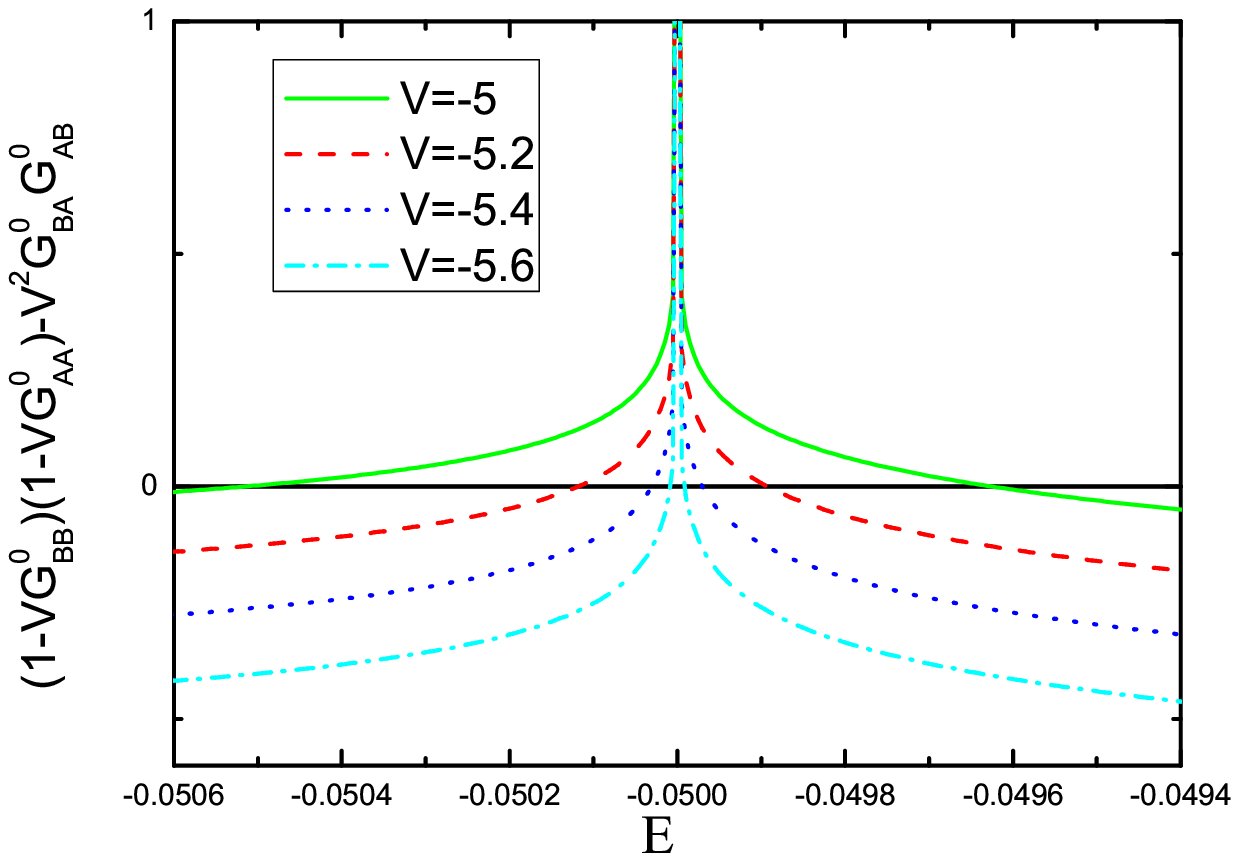} %
\includegraphics[height=3.0in,width=3.0in]{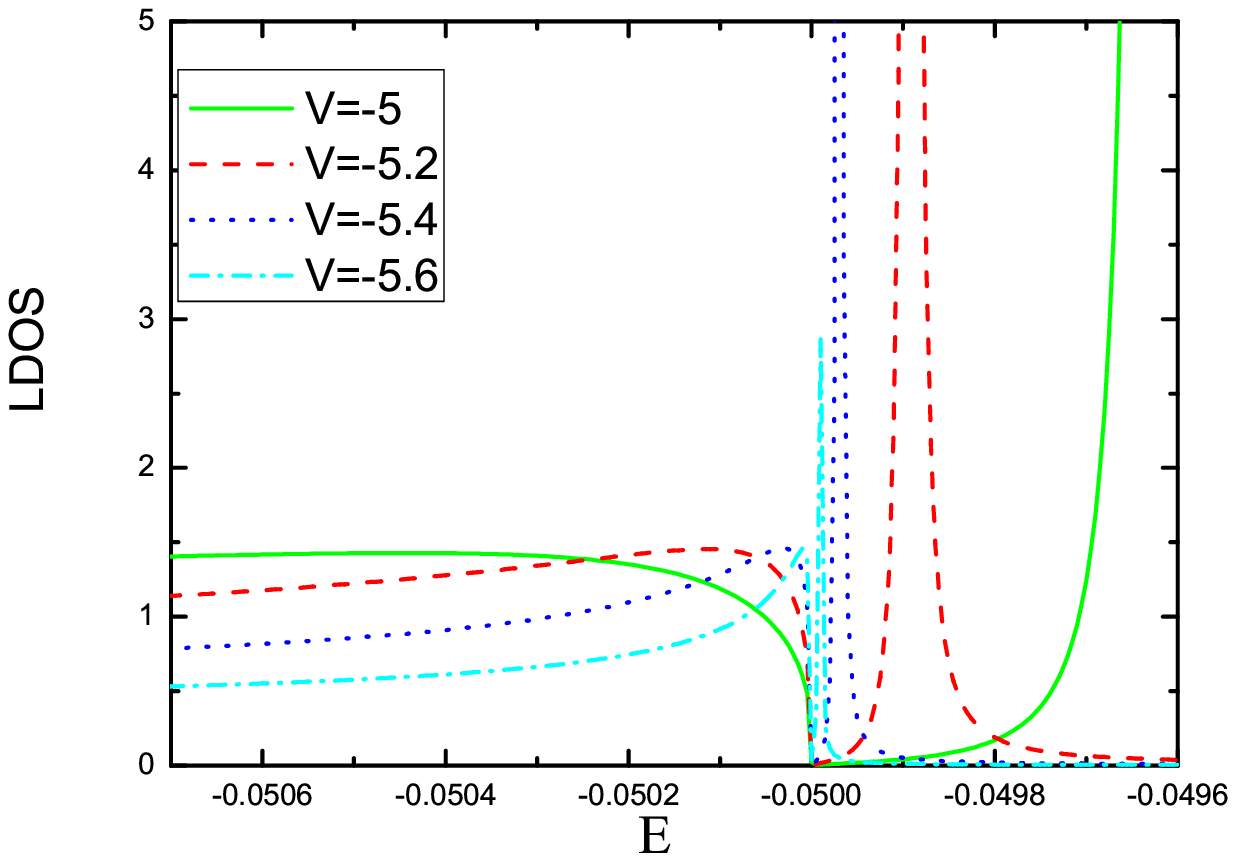}
\end{center}
\caption{ (color online) The behavior of LDOS at the valence band edge for
two impurities.}
\label{fig4}
\end{figure}

Then the Green function in the case $V_{2}=V$ is given by
\begin{equation}
G_{00}(\omega )=\frac{G_{00}^{0}(1-VG_{11}^{0})+V|G_{01}^{0}|^{2}}{%
(1-VG_{00}^{0})(1-VG_{11}^{0})-V^{2}|G_{01}^{0}|^{2}},  \label{lead_G_00_2}
\end{equation}%
Keeping the leading term for $G_{11}^{0}$, we find
\begin{equation}
G_{11}^{0}(\omega )\simeq -2mK\ln \omega ,
\end{equation}%
\begin{equation}
a_0=\frac{C\omega _{1}}{2mK},
\end{equation}%
where the quantity $C=|G_{01}^{0}|^{2}$ is finite near $\omega =-m$,
and we get the result near the bottom of the upper band:
\begin{equation}
a_0=2mK\omega _{2}\ln ^{2}\omega _{2},
\end{equation}%
The definitions of $\omega_{1},\omega_{2}$ are the same as in the case of
single impurity scattering discussed in the previous section. The conclusion is the
same: as the strength of the impurity interaction increases, the pole moves
towards the top of the bottom band, but never crosses it. Instead, the residue associated
with the pole decreases to zero. The key difference with the Coulomb case is that
these are short range impurities, and this leads to qualitatively different behaviour.



\subsection{Long-range asymptotes of the Green functions for a large number
of impurities}

In this subsection we generalize to some extent the results obtained in the
previous sections. We find the long-range asymptotic behavior of the Green function in
the case of multiple impurities located inside a finite area of the graphene
sheet. As a particular case this discussion includes the circular well,
discussed in the Dirac approximation, at the beginning of the article in Section II.
Knowing these Green functions' asymptotes we show that the spectral weight of
the state near the band edge (on the verge of entering into the lower
continuum) is zero and the screening charge is not significantly reshaped by
this state.\cite{Pereira08}

We rewrite Eq. (\ref{G_exp2.5}):
\begin{equation}
G_{\mathbf{R R}}= G^0_{\mathbf{R R}}+ \left[ \sum_{r \leq a} C_{\mathbf{r R}%
} G^0_{\mathbf{r R}} \right] {\large {/} \Delta}.
\label{4.1}
\end{equation}
Here we introduced the following notation: $R$ is the distance from
the center of the area in which the impurities are confined; we will
call this area the ``potential well"; \textbf{R} corresponds to the
site index outside the well; $a$ is the radius of the well; $r$ and
$r^{\prime}$ are the distances inside the circle of radius $a$, and
\textbf{r} is the index of the site inside the well.

The second term in Eq. (\ref{4.1}) represents the change induced by the impurity and is
responsible for the spectral weight of the bound state. Assuming $R \gg a$,
the following conclusions can be made about the second term. $\Delta$ does not
depend on $R$, but $C_{\mathbf{r R}}$ depends on $R$, and $C_{\mathbf{r R}} \sim
G^0_{\mathbf{{r} {R} }}$,  as the determinant $C_{\mathbf{r R}}$ contains
only one row with $G^0_{\mathbf{r R}}$.

Considering the sum over $\mathbf{r}$ in $\sum_{r \leq a}
C_{\mathbf{r R}} G^0_{\mathbf{r R}}$  we see that the $r$ dependency
comes only from the phase $\overrightarrow{k} \cdot
(\overrightarrow{R}- \overrightarrow{r})$ in the exponent of the
integrands (\ref{G_int1}, \ref{G_int2}). At the energies near the
band edge $\omega \rightarrow -m$, the part $\overrightarrow{k}
\cdot \overrightarrow{r}$ can be neglected, as the small $k$'s
produce most of the integral value and $\overrightarrow{k} \cdot
\overrightarrow{r}$ does not vary significantly near the Dirac
point. Therefore,
\begin{equation}
\sum_{r \leq a} C_{\mathbf{r R}} G^0_{\mathbf{r R}} \sim (G^0_{\mathbf{0 R}})^2.
\end{equation}
The spatial dependance of the second term in $G_{\mathbf{R R}}$ is
determined by the $(G^0_{\mathbf{0 R}})^2$, where $\mathbf{0}$
denotes some (arbitrary chosen) site in the impurity-occupied area.
To determine the asymptotic behaviour of $G_{\mathbf{0R}}^{0}$, we
use the method of stationary phase (see problem 5.2 in
\cite{Economou}), and get
\begin{equation}
G^{0}_{\mathbf{0 R}}(\omega \rightarrow -m)\sim \frac{\exp (-R\sqrt{\delta E}%
)}{(\delta E)^{1/4}\sqrt{R}},  \label{Green_asympt}
\end{equation}%
where $\delta E=m^{2}-\omega ^{2}$. Thus $G_{\mathbf{R R}}\sim \frac{\exp
(-2R\sqrt{\delta E})}{(\delta E)^{1/2}R}$, in qualitative agreement with the
asymptotic behavior predicted by the Dirac equation (\ref{solO_asymp}). As we can see
from the standard definition of the Green function \cite{Economou}:

\begin{equation}
G_{\mathbf{R R^{\prime }}}(\omega )=\sum_{n}\frac{\psi _{n}(\mathbf{R})\psi
_{n}(\mathbf{R^{\prime }})^{\ast }}{\omega -\omega _{n}}+\int dc\frac{\psi
_{c}(\mathbf{R})\psi _{c}(\mathbf{R^{\prime }})^{\ast }}{\omega -\omega _{c}}
\end{equation}
when $\omega \rightarrow \omega _{n}$ the $r$-dependency of $G_{\mathbf{R R}%
}(\omega )$ coincides with $\psi _{n}(\mathbf{R})\psi _{n}(\mathbf{R})^{\ast
}$ and we can judge if the state that is potentially crossing the band edge into the continuum is normalizable. In our case as
$\delta E\rightarrow 0$ the sum of $G_{\mathbf{R R}}$ over $\mathbf{R}$ in the
plane diverges, and so does the sum $\psi _{n}(\mathbf{R})\psi _{n}(\mathbf{R}%
)^{\ast }$ in the infinite lattice for any finite normalizing factor. Hence
the state merging into the continuum can be called non-normalizable or extended
and as such has zero spectral weight (in the thermodynamic limit). This confirms
our conclusion that there is no such phenomenon like supercritical screening
in the case of a localized potential (as opposed to a Coulomb potential) in graphene.

\section{Conclusions}

Our results are in agrement with common intuition developed in the physics of
shallow states in semiconductors. In particular it is not only a feature peculiar to gapped
graphene that properties of the states near the band edge are
strongly dependent on the long range ``tail" of the impurity potential.\cite{Pantelides}
In general, the effective mass approach (mostly determined by
long-range properties of a system) is in good agreement with exact
numerical methods for the energies near the band edge. It is exactly near the
band edge where the Coulomb tail becomes important while it is negligible in
computations related to deep levels.\cite{Rodriguez} As illustrated in Section II, in the continuum
limit a potential barrier emerges in the effective potential at large distances, due to the squaring of the Coulomb potential. Because of the two sublattices, a similar `squaring' occurs when the problem is solved
on a lattice, and lattice Green functions are utilized.\cite{Horiguchi}

\end{document}